\documentclass[longnamesfirst]{emulateapj}
\usepackage{apjfonts}
\usepackage{xspace}
\usepackage{epsfig}
\usepackage{amsmath}
\usepackage{graphicx}

\renewcommand\arcmin{\mbox{$^\prime$}\xspace}%
\renewcommand\arcsec{\mbox{$^{\prime\prime}$}\xspace}%
\renewcommand\micron{\mbox{$\mu$m}\xspace}%

\newcommand{\ergcms}{ergs~cm$^{-2}$~s$^{-1}$\xspace}

\newcommand{\ecmaa}{ergs~cm${-2}$~\AA$^{-1}$\xspace}

\newcommand{\msun}{$M_\sun$\xspace}

\newcommand{\panp}[1]{({\em #1})\xspace}
\newcommand{\pant}[1]{{\em #1}\xspace}
\renewcommand{\micron}{$\mu$m\xspace}

\newcommand{\HST}{{\em HST}\xspace}

\slugcomment{Accepted 2005 Aug 30 by the Astrophysical Journal Letters}
\shorttitle{Light Echo from SN~2003gd}
\shortauthors{Sugerman}

\begin{document}

\title{Discovery of a Light Echo from Supernova 2003gd}

\author{Ben E.K.~Sugerman}
\affil{Space Telescope Science Institute, 3700 San Martin Dr., Baltimore,
 MD 21286 U.S.A.}
\email{sugerman@stsci.edu}

\begin{abstract}
Archival {\em HST}/ACS data reveal details of a light echo around
SN~2003gd in the galaxy M74, only the fifth supernova around which
resolved echoes have been reported.  An echo is detected 0\farcs3 from
the supernova between PA 250\degr--360\degr, with fainter signal
present at a few other position angles.  This material lies $\sim$180
pc in front of the supernova, with a thickness of 60--120 pc, and may
delineate the disk of M74.  This structure has a gas density of
1--2~cm$^{-3}$, typical of the interstellar medium, however, the dust
grains are smaller than those found in our Galaxy, with maximum grain
sizes around 0.25\micron.  
Since only one epoch of data exists, in two wavebands and with
low signal-to-noise, deeper, annual visits should be made with {\em
HST} and ground-based adaptive optics telescopes.
\end{abstract}


\section{Introduction}

Supernova (SN) 2003gd was discovered by \citet{Eva03} on 2003 June
12.8 in a southern spiral arm roughly 151\arcmin from the nucleus of
the nearby galaxy NGC 628 (M74).  Its progenitor has been identified
by \citet{VDy03} and \citet{Sma04} as an $8^{+4}_{-2}$\msun red
supergiant.  Its photometric and spectral evolution are
very similar to the type II-plateau (II-P) SN 1999em
\citep[][, hereafter H05]{Hen05}, from which \citet{VDy03}
identify SN 2003gd as a type II-P, while H05 have estimated
the reddening as $E(B-V)=0.14 \pm 0.06$ \citep[of which
0.07 mags are Galactic,][]{FSD98}, and the distance to M74 as
$9.3 \pm 1.8$ Mpc.  Both groups conclude the SN was discovered roughly
87 days after explosion, which therefore occured around 2004 Mar 17.

Since the SN exploded near a spiral arm in M74, it is expected to
illuminate surrounding material in the form of scattered-light echoes.
Light echoes offer one of the most effective means to probe
circumstellar and interstellar structure.  As an example, echoes from
SN~1987A have been used to map the surrounding circumstellar
\citep{Sug05b} and interstellar \citep{Xu95} media, thereby probing
the progentor's mass-loss history, its location within its galaxy, and
the structure and history of the associated stars and gas.
For recent reviews on light echoes, see \citet{Sug03} and \citet{Pat05}.

To date, resolved echoes have been reported around only four
supernovae: SN~1987A \citep{cro88a}, SN~1991T \citep{sch94,spa99},
SN~1993J \citep{SC02,Liu03}, and SN~1998bu \citep{cap01}.  This {\em
Letter} presents the discovery and preliminary geometric 
and compositional analyses of a new light echo around SN 2003gd.

\section{Observations and Reductions}

Three epochs of publically-available data were accessed in the {\em
Hubble Space Telescope} (\HST) archive. The progenitor of SN 2003gd
was serendipitously observed on the Wide Field 2 chip of the Wide
Field and Planetary Camera 2 (WFPC2) in the F606W filter on 2002 Aug
25 and 28, with total exposure time of 3100 sec.  The SN was observed
with the Advanced Camera for Surveys (ACS) High Resolution Camera
(HRC) on 2003 Aug 1 (day 137 after outburst), and again on 2004 Dec 8
(day 632).  The SN is too bright in the former epoch to be reliably
PSF subtracted, thus these earlier data were of limited use.  In 2004,
the SN was imaged in the F435W and F625W filters, with total
integration times of 840 and 350 sec, respectively.

Pipeline-calibrated data were first filtered to mask warm pixels by
rejecting single pixels that deviated by more than 3-$\sigma$ from the
variance of the other 8 pixels in a 9-pixel moving box.  Individual
exposures were combined using the {\tt multidrizzle} task within {\tt
stsdas} to reject cosmic rays and perform geometric-distortion
corrections.  The WFPC2 data were drizzled with a final pixel scale of
0\farcs065~pix$^{-1}$, to increase the resolution of the PSF.

To intercompare images, HRC data were geometrically-registered to a
common orientation using a shift, rotation, and scaling.  The F625W
HRC image from 2004 was drizzled to 0\farcs065~pix$^{-1}$ and
registered to the WFPC2 field using a second-order polynomial.  All
registrations had residuals $<0.1$ pixel rms.  HRC data, and the WFPC2
F606W and HRC F625W images, were PSF-matched, photometrically scaled,
and subtracted using the {\tt difimphot} image-subtraction techniques
of \citet{TC96}, as described in \citet{Sug05a}, using Tiny Tim model
PSFs that underwent identical geometric transformations.  

Photometric calibrations were adopted from the image headers, and
unless otherwise noted, magnitudes are expressed in STMAGS.
Conversion to Johnson/Cousins magnitudes was accomplished as specified
in \citet{Sir05}.  For example, PSF-fitted photometry of the SN in
2004
yields $m_{F435W}=23.39\pm0.13$ and $m_{F625W}=23.67\pm0.12$, which
translate into $B=23.98\pm0.15$ and $R=23.13\pm0.16$.  Note that the
transformations, while color dependent, are determined from stellar
sources, and since they do not account for the strong emission lines
present in a SN spectrum, the uncertainties are likely to be larger
than the formal errors quoted.

\section{Echo Identification and Geometry}

\begin{figure*}\centering
\includegraphics[height=6in,angle=-90]{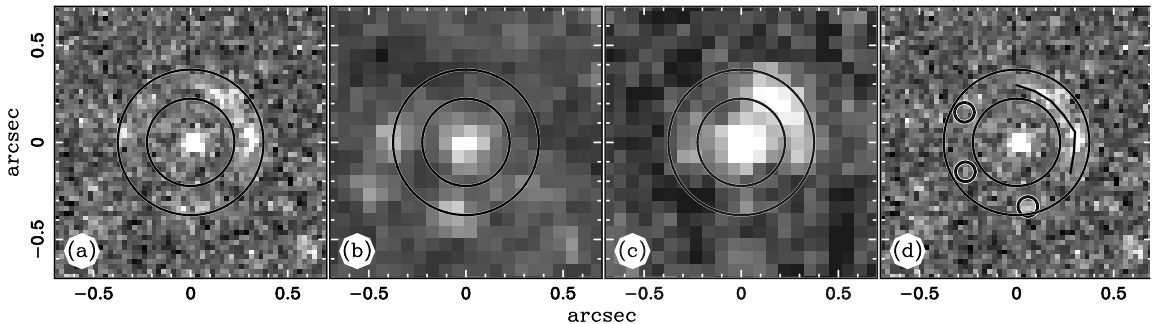}
\caption{\HST images of a 1\farcs4 field surrounding SN 2003gd. North
  is up and east is left.  The SN (or progenitor) is at the center of
  each frame.  Circles of radii 0\farcs225 and 0\farcs375 roughly
  delimit the radial extent of the echoes.  \panp{a} ACS/HRC F435W
  image from 2004.  \panp{b} WFPC2 F606W image of the progenitor,
  drizzled to 0\farcs065~pix$^{-1}$.  \panp{c} PSF-matched difference
  image between the ACS/HRC F625W data from 2004 drizzled to
  0\farcs065~pix$^{-1}$, and the data in panel \pant{b}.  \panp{d}
  Panel \pant{a} with the neighboring stars removed.  The bright echo
  loci are marked with a black line, while fainter echo positions are
  denoted with small black circles.
 \label{2003gdtile}}
\end{figure*}

Figure \ref{2003gdtile}\pant{a} shows the ACS/HRC F435W image from
2004.  An arc of flux is present between 0\farcs23 and 0\farcs38 from
the SN, between PA 260\degr--340\degr, as well as a few isolated,
point-source like features at other position angles.  A comparison
with the WFPC2 F606W image from 2002 (panel \pant{b}) shows that two
of these sources are nearby stars, yet there is no observable arc of
flux northwest of the SN.

The exposure time of the F625W HRC image from 2004 is
short compared to the F606W WFPC2 integration, making the HRC image
very noisy at its native resolution.  However, once resampled to the
WFPC2 resolution, this image was succesfully PSF-matched to and
differenced from the WFPC2 data, as shown in panel
\panp{c}.  The nearby stars have cleanly subtracted away, revealing
the bright arc to the northwest, and potential sources at PAs
60\degr, 100\degr, and 190\degr.  The bright arc is consistent with a
scattered-light echo, while the nature of these isolated
sources (i.e.,\ echoes from more localized density
enhancements versus detector artifacts) is unclear.  

The positions of the three companion stars were measured on the WFPC2
frame and transformed to the ACS coordinate system, then subtracted
away from the 2004 data using Tiny Tim model PSFs and the {\tt
daophot} package.  The result in F435W is shown in panel \panp{d}.  To
measure the positions of the echoes, and to search for faint flux not
readily apparent to the eye, radial profiles of angular width 20\degr
were inspected, and peaks were fit with moffat profiles to find the
center and width of any extended features \citep{Sug05a}.  All
identified echo  sources have been marked in panel \panp{d}.  The
bright arc extends from 0\farcs28 to 0\farcs32, with coherent signal
detected from PA 250\degr--360\degr.  Adopting a distance of 9.3 Mpc,
these correspond to separations on the sky of $\rho=$41--47
lt-yr\footnote{With units of years and light-years, the echo equations
simplify considerably since $c=1$, hence the somewhat unorthodox use
of distances in lt-yr throughout this paper.}.

\begin{figure}\centering
\includegraphics[height=3.25in,angle=-90]{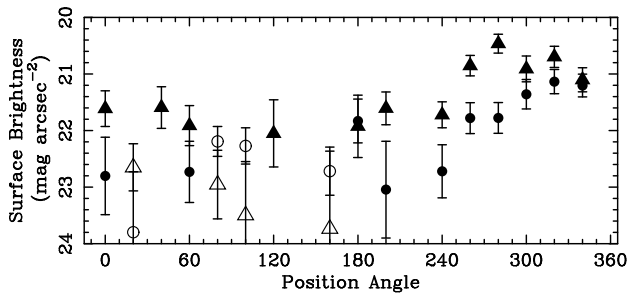}
\caption{Surface brightness measured in 20\degr wedges in PA from
$\rho=$0\farcs23 to 0\farcs38 in F435W (triangles) and F625W
  (circles).  Filled symbols denote position angles at which echoes
  were identified.
\label{plsb}}
\end{figure}

The surface brightnesses of the echoes as measured in annuli with
20\degr arclength are plotted in Fig.\ \ref{plsb}.  For echoes already
identified above, the moffat-profile width was used as the annular
width, while at all other PAs the annuli were at 0\farcs28 and
0\farcs32.  In general, the fluxes of identified echoes track each
other across PA, however there is considerable scatter due to the low
signal-to-noise of these short observations.  Echoes between PA
0\degr--240\degr are considered candidates at this time since their
detections are only marginal; in some cases the echo was not even
detected in F625W.

As noted in \S1, SN 2003gd was discovered $\sim$87 days
after explosion.  However, H05 have argued it is sufficiently
similar to SN 1999em to warrant using the
latter's photometry and spectra as surrogates at earlier times.  Filling
in the first $\sim80$ days of photometry with that of SN 1999em
(H05 and references therein), the SN 2003gd light curves have roughly 120
day plateaus, peaking very early in blue and around 60 days in the
red.  Unfortunately, ACS spatial
resolution cannot differentiate the temporal color peaks in these echoes,
as seen in echoes around N V838 Mon \citep{Bon03}.

If we assume maximum light occured around 60 days, the line-of-sight
distance of each echo is computed as $z=\rho^2/(2ct)-ct/2$,
where $t$ is the time since maximum light.  The center of the bright
arc therefore lies between 530 and 650 lt-yr in front of the SN, while
the other possible echoes extend up to $z=$780 lt-yr.  Using the naming
scheme of \citet{Xu95}, we label the bright arc NW600.

\begin{figure*}\centering
\includegraphics[width=5in,angle=0]{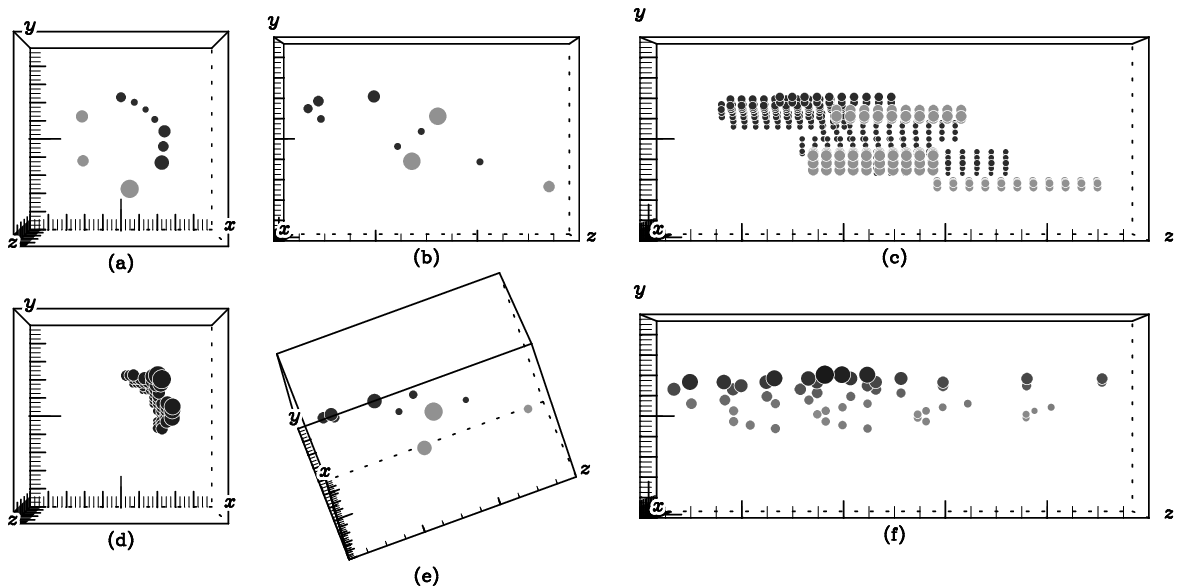}
\caption{Three-dimensional renderings of the observed echo locations.
  Axes are defined with $x$ increasing west, $y$ increasing north, and
  $z$ increasing toward the observer.  Axis labels denote the
  direction of increasing value.  Along the $x$ and $y$ axes, major
  ticks mark 20 lt-yr, while in $z$ major ticks mark 100 lt-yr; the
  longest ticks mark the origin. The coordinates and axes have been
  given a slight perspective transformation, and larger point size
  indicates closer position to the view. Dark points denote the bright
  echo arc (NW600) from PA 250\degr--360\degr, with the other echoes
  marked in light grey. The $z$ axes run from 500 to 800 lt-yr in
  panels \panp{b}--\panp{c}, and from 400--900 lt-yr in panels
  \panp{d}--\panp{e}.
  \panp{a} Face-on view as observed on the sky.  These data are also
  shown \panp{b} from the side (rotated 90\degr) with the observer to
  the right, and \panp{d} from an oblique view, in which echoes appear
  to lie roughly along a plane.
  \panp{c} As panel \pant{b}, but showing the spatial extent of dust
  along the line of sight ($z$) and in position angle that comprises
  each individual echo position.
 All pixels from the 2004 ACS data belonging to NW600 are rendered
 \panp{d} face on, and \panp{f} from the side.  In panel \panp{f},
 points are shaded using simple ray tracing assuming a single light
 source behind the reader.
 \label{3dpl}}
\end{figure*}

The three-dimensional positions of the echoes have been rendered in
Figure \ref{3dpl} using the techniques described in \citet{Sug05a} and
briefly annotated in the figure caption.  Panel \panp{a} shows the
positions from Fig.\ \ref{2003gdtile}\pant{d} as viewed on the plane
of the sky.  From the side (panel \pant{b}), the echo points appear to
lie along a steeply-inclined plane, with its normal inclined roughly
60\degr to the line of sight at PA 340\degr.  This is shown in the
oblique view of panel \panp{e}.

As shown in Fig.\ 2 of \citet{Sug03}, an echo has a width $\Delta\rho$
resulting from a convolution of the depth of the material $\Delta z$
and the finite duration of the light pulse $\Delta t$.  Using the
widths measured during radial-profile fitting, and Eq.\ (11) of
\citet{Sug03}, the depths of the echoes are 110--160 lt-yr.  The
spatial extent of each echo measurement has been shown in Fig.\
\ref{3dpl}\pant{c}, where each point in the panel \panp{b} is
replicated through its range of $\Delta z$ and position angle.

While panel \panp{c} is still suggestive of a highly-inclined plane,
the case for this structure becomes less compelling when the light
grey points (echo {\em candidates}) are ignored.  Consider also Fig.\
\ref{3dpl}\pant{d} and \pant{f}, which show the three-dimensional
positions of all image pixels that are part of the northwestern arc.
From the side, this locus appears consistent with a thick ($\Delta
z=300-400$ lt-yr) sheet of material that is roughly perpendicular to
the line of sight.  This underscores the fundamental problem that it
is very difficult to identify the geometry of an echoing structure
with a single epoch of observation.

\section{Dust Analysis}

When echoes are observed in multiple wavebands, characteristics of the
scattering dust (i.e.\ dust composition, grain size, and density) can
be constrainted by comparing the SN and echo photometry, using e.g.\
the formalisms presented in \citet{Sug03}.  Since the SN 2003gd echoes
were observed in only two filters, and since the signal-to-noise of
the observations is low, such analyses have limited accuracy.  In
particular, with only two wavebands, a variety of dust mixtures may
fit the data equally well \citep[see][]{Pat05}.  As such, only the
{\em total} surface brightnesses of the NW600 echo are considered,
which are $\mu_{F435W}=20.8\pm 0.2$ and $\mu_{F625W}=21.4\pm 0.3$ mags
arcsec$^{-1}$.

The light curves of SN 1999em and 2003gd (H05 and references therein)
integrated over the first 140 days yield $B$, $V$, and $R$ fluences of
$(7.0, 8.0, 7.0)\times10^{-8}$ \ecmaa,
respectively.  When integrated over the first 140 days, the spectra of
SN 1999em\footnote{Acquired from the SUSPECT supernova spectrum
archive, http://suspect.nhn.ou.edu.}  corrected for extinction toward
SN 2003gd
match these to better than 1\%.  yielding fluences of
$6.5\times10^{-8}$ and $7.5\times10^{-8}$ \ecmaa in F435W and F625W.

Comparison of the echo surface brightness to the SN fluence yields a
color shift of $\Delta(m_{F435W}-m_{F625W})=-0.8 \pm 0.4$.  Two
extremal color shifts are for Galactic \citep{WD01} and
Rayleighan\footnote{Only small-particles, hence scattering efficiency
$\propto\lambda^{-4}$} dust, which using the scattering function
$S(\lambda,\mu)$ of \citet{Sug03} yields
$\Delta(m_{F435W}-m_{F625W})=-0.3$ and $-1.2$, respectively.  This
rules out purely Rayleighan particles, suggesting instead that the
dust is similar to Galactic dust, but with smaller maximum grain
sizes.

\begin{deluxetable}{lccccccc}
\tablecaption{Dust modeling results \label{tbl2}}
\tablewidth{0pt}
\tablehead{
 \colhead{Model} & \colhead{$a_{max}$} &  
  \multicolumn{3}{c}{$n_{\rm H}$ (cm$^{-3}$)}  & 
  \multicolumn{3}{c}{$A_V/\Delta z_{100}$\tablenotemark{a}} \\
 \colhead{} & \colhead{\micron} &
 \colhead{C} & \colhead {Si} & \colhead{C+Si} &
 \colhead{C} & \colhead {Si} & \colhead{C+Si} 
}
\startdata
 WD01\tablenotemark{a} & 0.25 & ... & 1.3 & 1.1 & ... & 0.033 & 0.050 \\
                & 0.13 &  30.   & ... & ...& 0.38 & ... &... \\
 MRN\tablenotemark{b}  & 0.25 & ... & 2.3 & 1.5 & ... & 0.037 &  0.083 \\
                & 0.13 &  15.   &...  &... & 0.40  &...  &... \\
\enddata
\tablenotetext{a}{mags per 100 lt-yr of material in $z$.}
\tablenotetext{b}{\citet{WD01} Galactic dust with $R_V=3.1$ and $b_C=6\times10^{-5}$.}
\tablenotetext{c}{\citet{MRN77} dust with $n(a)\propto a^{-3.5}$}
\end{deluxetable}

A more detailed comparison of the echo surface brightness to a range
of dust models \citep{Sug03} yields the results summarized in Table 1.
Each model used the dust properties in the first column, with minimum
grain size $a_{min}=5\times10^{-4}$\micron, and maximum grain size
given in column 2.  The associated gas density $n_{\rm H}$ and
resulting extinction are given for purely carbonaceous (C), purely
silicate (Si), and a Galactic mixture (C+Si) of dust. A Galactic dust
composition with $a_{max}\sim 0.25$\micron and $n_{\rm
H}\sim$1--2~cm$^{-3}$ is a good fit to the echoes, while
slightly-higher densities are required for Si dust. By contrast, C
dust requires smaller grains and at least an order of magnitude more
dust.  

An additional constraint is provided by the reddening H05 measure to
the SN (\S1), which for $R_V=3.1$ and with 0.07 mag of Galactic
extinction removed, yields $A_V=0.36$.  C dust is only realistic if
the echoes occur in a thin sheet of material.  Similarly, Si dust
requires a sheet at least twice as thick as actually observed.
Assuming the scattering occurs in a sheet roughly 300-400 lt-yr thick,
the echoes are reasonably-well fit by the Galactic composition.
However, a better fit may require a higher proportion of small carbon
grains than are found in Galactic dust.  For example, 
75\% carbon grains with $a_{max}$=0.15\micron and 25\%
silicate grains with $a_{max}=$0.25\micron yield a gas density of
2.2~cm$^{-3}$ and $A_V/\Delta z_{100}$=0.078, which satisfies the
observed extinction for a slab of dust 400 lt-yr thick.  Again, since
the echoes have only been measured at two wavebands, these specific
results are far from conclusive.

\section{Discussion}

The disk of M74 is inclined $8\degr\pm 5\degr$ at PA $23\degr\pm
6\degr$ \citep{GGA91} with a scale height of $730\pm 200$ lt-yr
\citep{Pen88}.  As discussed in \S3, there are two hypotheses for the
geometry of the echoing dust.  The thick slab with shallow inclination
is consistent with the geometry of the disk, and a density of 1--2
cm$^{-3}$ is consistent with ambient density of the Galactic
interstellar medium.  In this case, the SN has illuminated the disk of
M74, and is itself located behind the disk.  On the other hand, a
thinner, highly-inclined plane of dust might delineate the shell of a
(remnant) superbubble blown from a nearby OB association of stars
\citep[See for example][]{Xu95}.  A blue cluster of stars with
integrated color of $B-V=-0.25$
is located about 3\arcsec from the SN at PA 65\degr.  Alternatively,
there are at least two additional blue clusters of stars within 12\arcsec
of the SN, all of which may have formed a complex network of bubbles
in front of (or around) SN 2003gd.  Obviously, additional observations
are required to properly understand the geometry of this material.

Adopting the thick-sheet model, this echo will expand at $d\rho/dt =
c(z+ct)/\rho =$0\farcs1~yr$^{-1}$.  Thus, observations should continue
at yearly intervals.  The observed echo is only a $\sim$90\degr arc,
with scattered hints of echoes that form a more complete loop
around the SN.  Since the exposure times of the 2004 ACS data were
short, it is unclear whether fainter echo flux is present at all
position angles.  Future integrations should probe at least one
magnitude deeper, and should include the complete $U-I$ range
to properly study the dust characteristics.

The total echo fluxes are $4.0\times10^{-16}$ and $2.6\times10^{-16}$
\ergcms in the F435W and F625W filters, respectively, corresponding to
$B=24.2\pm0.1$ and $R=23.9\pm0.1$ mag.  Using the same scattering
codes as above, these yield a $V$-band magnitude of $24.2\pm0.2$.
Around day 137 (the 2003 August ACS observations), the echo was at
$\rho=0\farcs1$ but only 2\% brighter.  Compared to the SN fluxes of
19.1, 17.4, and 16.3 in $B$, $V$, and $R$ respectively (H05), the echo
was at least 5 mag fainter and therefore lost within the wings of the
possibly-saturated PSF.  However, a a small correction must be made to
earlier photometry to remove any echo contributions.

As a final note, H05 report photometry of the SN from day 493 as
$B=21.76\pm0.06$ and $R=20.59\pm0.09$ mags.
They further speculate that dust may be forming in the ejecta of the
SN, resulting from a small blueshift in H$\alpha$ emission, and a
marginally-significant drop in this late-time photometry.
Following \citet{Tur90}, the average Type IIP decay rates for these
filters are are roughly 0.7 and 1.0 mags per 100 days, yielding
expected $B$ and $R$ values of 22.7 and 22.0.  Compared to those
measured on day 632, the photometry has decreased about 1.3 and 0.9
mags in $B$ and $R$, consistent with an increase in the extinction
toward the SN of  $\Delta A_V=1$ mag, and further consistent with dust
formation at these late times \citep[see also][]{Woo93}.

\acknowledgments Sincere thanks to Steve Lawrence and the referee,
 Fernando Patat, for their careful reading of this
 manuscript.  This work was based on observations made with the
 NASA/ESA Hubble Space Telescope, obtained from the Data Archive at
 the Space Telescope Science Institute, which is operated by the
 Association of Universities for Research in Astronomy, Inc., under
 NASA contract NAS 5-26555.  This research was supported by STScI
 grants 10204 and 82301.

\end{document}